\documentclass[twocolumn,aps,prd,showpacs,10pt,superscriptaddress]{revtex4-1}
\usepackage{amssymb}
\usepackage{amsmath}
\newcommand{\vect}[1]{\boldsymbol{#1}} 
\usepackage{multirow}
\usepackage{graphicx}
\usepackage[usenames,dvipsnames,svgnames]{xcolor}
\usepackage{hyperref}
\hypersetup{
pdfnewwindow=true,      
colorlinks=true,        
linkcolor=Blue,         
citecolor=Blue,         
filecolor=Blue,         
urlcolor=Blue           
}

\begin{document}

\title{Thermal Transport in MoS$_2$ from Molecular Dynamics using Different Empirical Potentials}
\author{Ke Xu}
\affiliation{Jiangsu Key Laboratory of Advanced Food Manufacturing Equipment and Technology, Jiangnan University, 214122 Wuxi, China}
\author{Alexander J. Gabourie}
\email{gabourie@stanford.edu}
\affiliation{Department of Electrical Engineering, Stanford University, Stanford, CA 94305, USA}
\author{Arsalan Hashemi}
\email{arsalan.hashemi@aalto.fi}
\affiliation{Department of Applied Physics, Aalto University, FI-00076 Aalto, Finland}
\author{Zheyong Fan}
\email{brucenju@gmail.com}
\affiliation{School of Mathematics and Physics, Bohai University, Jinzhou, China}
\affiliation{QTF Centre of Excellence, Department of Applied Physics, Aalto University, FI-00076 Aalto, Finland}
\author{Ning Wei}
\email{weining@mail.tsinghua.edu.cn}
\affiliation{Jiangsu Key Laboratory of Advanced Food Manufacturing Equipment and Technology, Jiangnan University, 214122 Wuxi, China}
\author{Amir Barati Farimani}
\affiliation{Mechanical Engineering Department, Carnegie Mellon University,5000 Forbes Avenue, Scaife Hall 310,Pittsburgh, PA 15213 }
\author{Hannu-Pekka Komsa}
\affiliation{Department of Applied Physics, Aalto University, FI-00076 Aalto, Finland}
\author{Arkady V. Krasheninnikov}
\affiliation{Department of Applied Physics, Aalto University, FI-00076 Aalto, Finland}
\affiliation{Institute of Ion Beam Physics and Materials Research, Helmholtz-Zentrum Dresden-Rossendorf, 01328 Dresden, Germany}
\author{Eric Pop}
\affiliation{Department of Electrical Engineering, Stanford University, Stanford, CA 94305, USA}
\author{Tapio Ala-Nissila}
\affiliation{QTF Centre of Excellence, Department of Applied Physics, Aalto University, FI-00076 Aalto, Finland}
\affiliation{Center for Interdisciplinary Mathematical Modeling and Department of Mathematical Sciences, Loughborough University, Loughborough, Leicestershire LE11 3TU, UK}

\date{\today}

\begin{abstract}
Thermal properties of molybdenum disulfide (MoS$_2$) have recently attracted attention related to fundamentals of heat propagation in strongly anisotropic materials, and in the context of potential applications to optoelectronics and thermoelectrics. Multiple empirical potentials have been developed for classical molecular dynamics (MD) simulations of this material, but it has been unclear which provides the most realistic results. Here, we calculate lattice thermal conductivity of single- and multi-layer pristine MoS$_2$ by employing three different thermal transport MD methods: equilibrium, nonequilibrium, and homogeneous nonequilibrium ones. These methods allow us to verify the consistency of our results and also facilitate comparisons with previous works, where different schemes have been adopted. Our results using variants of the Stillinger-Weber potential are at odds with some previous ones and we analyze the possible origins of the discrepancies in detail. We show that, among the potentials considered here, the reactive empirical bond order (REBO) potential gives the most reasonable predictions of thermal transport properties as compared to experimental data. With the REBO potential, we further find that isotope scattering has only a small effect on thermal conduction in MoS$_2$ and the in-plane thermal conductivity decreases with increasing layer number and saturates beyond about three layers.  We identify the REBO potential as a transferable empirical potential for MD simulations of MoS$_2$ which can be used to study thermal transport properties in more complicated situations such as in systems containing defects or engineered nanoscale features. This work establishes a firm foundation for understanding heat transport properties of MoS$_2$ using MD simulations.

\end{abstract}

\maketitle
\section{Introduction}
Atomically thin molybdenum disulfide (MoS$_2$) is a layered material which has attracted enormous interest due to its electronic and optical properties \cite{mak2010prl,splendiani2010nl,wang2012nnano}. In electronic device applications such as transistors based on MoS$_2$, device self-heating \cite{yalon2017nl} could limit the saturation velocity, which ultimately limits device performance \cite{smithe2018nl}. On the other hand, the large, tunable Seebeck coefficient (thermopower) \cite{pu2016prb} and power factor \cite{hippalgaonkar2017prb} of MoS$_2$ make it a promising candidate for thermoelectric applications. Knowledge of the thermal transport properties of MoS$_2$ is crucial in both types of applications.

There have been several studies of thermal transport properties of MoS$_2$ both experimentally \cite{muratore2013apl,liu2014jap,zhu2016nc,jiang2017am,sahoo2013jpcc,jo2014apl,yan2014acsnano,zhang2015acsami,bae2017nanoscale,yarali2017afm,aiyiti2018nanoscale} and theoretically \cite{li2013apl,gu2016jap,lindroth2016prb,cepellotti2017nl,ding2015nt,jin2015sr,kandemir2016nt,hong2016jpcc}. Experimentally, the measured in-plane thermal conductivity values in bulk natural crystal are about 100 W m$^{-1}$ K$^{-1}$ \cite{liu2014jap,zhu2016nc,jiang2017am}, while those in exfoliated or synthesized single- and multi-layer MoS$_2$ are typically lower \cite{sahoo2013jpcc,jo2014apl,yan2014acsnano,zhang2015acsami,bae2017nanoscale,yarali2017afm,aiyiti2018nanoscale}, varying from $13.3 \pm 1.4$ to $ 84 \pm 17$ W m$^{-1}$ K$^{-1}$. The measured through-plane thermal conductivity values in bulk MoS$_2$ are more than one order of magnitude smaller \cite{muratore2013apl,liu2014jap,zhu2016nc,jiang2017am}, ranging from $2.0\pm 0.3$ to $4.75 \pm 0.32$ W m$^{-1}$ K$^{-1}$.  Theoretically, Li \textit{et al.} \cite{li2013apl} first calculated the in-plane thermal conductivity of single-layer MoS$_2$ using the Boltzmann transport equation (BTE), with the third-order anharmonic force constants obtained from quantum mechanical density functional theory (DFT) calculations. The thermal conductivity of naturally occurring MoS$_2$ was calculated to be about $108$ W m$^{-1}$ K$^{-1}$ for a $10$-$\mu$m-long sample \cite{li2013apl}. Using similar method, Gu \textit{et al.} \cite{gu2016jap} found that the in-plane thermal conductivity of $10$-$\mu$m-long samples in layered, naturally occurring MoS$_2$ monotonically reduces from $138$  W m$^{-1}$ K$^{-1}$ to $98$  W m$^{-1}$ K$^{-1}$ when the thickness increases from one to three layers. 

While the BTE approach is widely used in predicting the thermal conductivity of materials, and a handful of computer codes \cite{li2014cpc,chernatynskiy2015cpc,togo2015prb,carrete2017cpc,tadano2014jpcm} are available for the calculations, the method has its limitations. First, it is based on perturbation theory and it is usually assumed that fourth- and higher-oder phonon-phonon interactions are unimportant, which is not valid at high temperatures. Second, since the computational cost of the BTE approach increases rapidly with the supercell size, it is impractical for studying spatially complex structures such as those with defects, grain boundaries or engineered nanostructures. 

The above limitations for the BTE approach can be overcome by classical molecular dynamics (MD) methods, which are nonperturbative and scale linearly with the simulation cell size. Nevertheless, predictions from classical MD simulations are sensitive to the empirical potential used. A few works \cite{ding2015nt,jin2015sr,kandemir2016nt,hong2016jpcc} have employed MD simulations to study heat transport in suspended MoS$_2$, using the Stillinger-Weber (SW) potential  \cite{stillinger1985prb} modified and parameterized by Jiang  \textit{et al.} \cite{jiang2013jap} or Kandemir \textit{et al.} \cite{kandemir2016nt}. While many insights have been gained from previous MD simulations \cite{ding2015nt,jin2015sr,kandemir2016nt,hong2016jpcc}, there is an apparent inconsistency between two types of thermal conductivity calculations, namely the equilibrium (Green-Kubo) and nonequilibrium methods, that has not been resolved. Using nonequilibrium MD simulations and the potential by Jiang \textit{et al.} \cite{jiang2013jap}, Ding \textit{et al.} \cite{ding2015nt} obtained an in-plane thermal conductivity of $\kappa=19.76$ W m$^{-1}$ K$^{-1}$ for pristine single-layer MoS$_2$, but a very different value of $\kappa=116.8$ W m$^{-1}$ K$^{-1}$ has been obtained by Jin \textit{et al.} \cite{jin2015sr} using the equilibrium method with the same potential. Because the equivalence between the equilibrium and the nonequilibrium MD methods is well established both theoretically and in properly executed MD simulations \cite{dong2018prb}, it is imperative to examine this inconsistency in detail for the present case. 

On the other hand, MoS$_2$ is widely used as a solid lubricant \cite{winer1967wear} and a sophisticated empirical potential based on the proven framework of the Abell-Tersoff-Brenner potentials \cite{abell1985prb,tersoff1988prb,brenner1990prb} has already been developed by Liang \textit{et al.} \cite{liang2009prb,liang2012prb} to simulate friction between MoS$_2$ layers. To our knowledge the potential by Liang \textit{et al.} has not been used for heat transport applications and there is, so far, no detailed comparison between the above-mentioned potentials regarding thermal transport. In view of the importance of the quality of the empirical potential in MD simulations, it is of great interest to evaluate these potentials through a careful comparison of the simulation results to the available experimental data. 

To this end, we study here heat transport in single-layer, multi-layer, and bulk MoS$_2$ using extensive MD simulations. First, we examine the consistency between various MD based methods for heat transport and compare our results closely with previous works \cite{ding2015nt,jin2015sr,kandemir2016nt,hong2016jpcc}. Then, we benchmark our MD results against the available experimental data \cite{muratore2013apl,liu2014jap,zhu2016nc,jiang2017am,sahoo2013jpcc,jo2014apl,yan2014acsnano,zhang2015acsami,bae2017nanoscale,yarali2017afm,aiyiti2018nanoscale}, evaluating the performance of the empirical potentials we considered and rationalizing the theoretical and experimental results. This work establishes a firm foundation for understanding heat transport properties of MoS$_2$ using MD simulations.

This paper is organized as follows: Section \ref{section:model} introduces the simulation model of this work, which we base the discussion of the various empirical potentials on in Sec. \ref{section:potential}. In Sec. \ref{section:method}, we introduce three different MD methods for thermal conductivity calculations. After presenting our thermal conductivity calculation results for single-layer MoS$_2$ in Sec. \ref{section:hnemd}, we make detailed comparisons with previous MD works in Sec. \ref{section:comparison}. In Sec. \ref{section:experiments}, we evaluate the performance of empirical potentials from various perspectives and compare our MD results to experimental data. Section \ref{section:conclusions} summarizes and concludes our work. 

\section{Models and Methods}

\subsection{Atomistic model of molybdenum disulfide \label{section:model}}

\begin{figure*}[htb]
\begin{center}
\includegraphics[width=1.8\columnwidth]{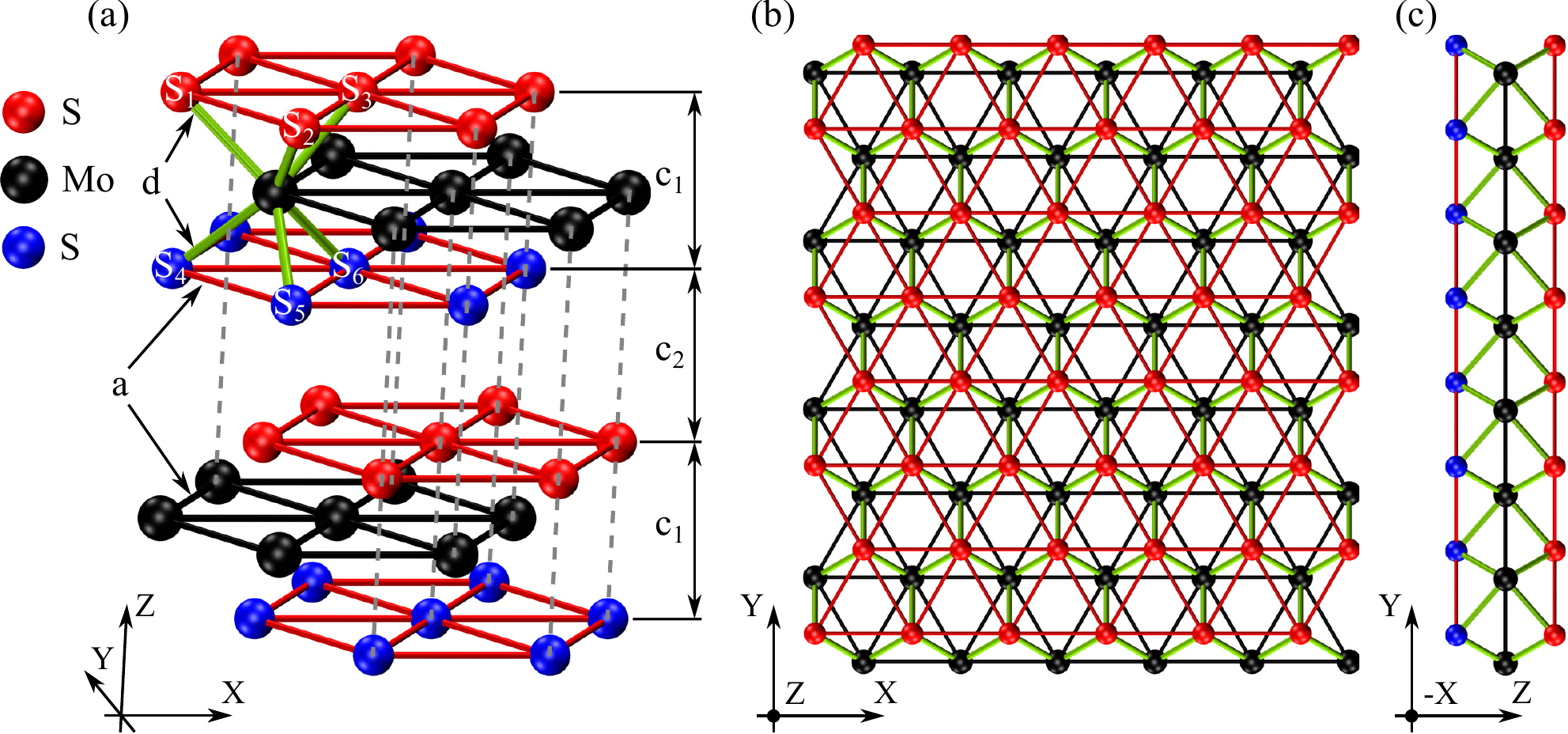}
\caption{A schematic illustration of the atomistic structure of multi-layer MoS$_2$ in the trigonal prismatic H phase.
For simplicity, a two-layer system is shown in (a). Each layer of MoS$_2$ consists of three sublayers, including a middle Mo sublayer (in black), a top S sublayer (in red), and a bottom S sublayer (in blue). The shortest Mo-Mo pairs (in black), S-S pairs (in red), and Mo-S bonds (in green) have equilibrium distances $a$, $a$, and $d$, respectively. The intralayer and interlayer distances of two S sublayers are shown by $c_1$ and $c_{2}$, receptively. (b) Single-layer MoS$_2$ viewed from the top. There is a zigzag-shaped edge in the $x$ direction and an armchair-shaped edge in the $y$ direction. (c) A side view for single-layer MoS$_2$. Experimental values are \cite{winer1967wear}: $a \approx 3.16$ \AA, $d \approx 2.42$ \AA, $c_1 \approx 3.24$ \AA, $c_2 \approx 2.90$ \AA.}
\label{f_model}
\end{center}
\end{figure*}

Figure \ref{f_model} shows the atomistic structure of a multi-layer MoS$_2$ system in the trigonal prismatic H phase with hexagonal symmetry. Each layer of MoS$_2$ consists of a Mo sublayer in the middle sandwiched by two S sublayers.
The in-plane lattice constant is \cite{winer1967wear} $a \approx 3.16$ \AA, which is the equilibrium nearest neighbor Mo-Mo (or S-S) distance. The intralayer distance of the S sublayers is $c_1 \approx 3.24$ \AA~ and the nearest interlayer distance of the S sublayers is $c_2 \approx 2.90$ \AA. In single layers, each Mo atom with trigonal prismatic coordination links to the six nearest S atoms.
The equilibrium Mo-S bond length is $d \approx 2.42$ \AA. For two-dimensional materials, the thickness must be specified to calculate the effective system volume needed for calculating the effective thermal conductivity in three dimensions. In this work, we follow the convention in literature to set the thickness of $n$-layer MoS$_2$ to $6.15\times n$ \AA.

\subsection{Empirical interatomic potentials for molybdenum disulfide\label{section:potential}}

We consider multiple, widely used  empirical interatomic potentials for MD simulations of MoS$_2$ \cite{liang2009prb,liang2012prb,stewart2013msmse,jiang2013jap,kandemir2016nt}. We implement all the potentials in the open source GPUMD package \cite{fan2013cpc,fan2017cpc,gpumd} and confirm that GPUMD and the LAMMPS package \cite{plimpton1995jcp,lammps} give consistent forces for all potentials. 

\subsubsection{REBO-LJ potential for Mo-S systems}

Liang \textit{et al.} \cite{liang2009prb,liang2012prb} developed a potential in 2009 combining a REBO (reactive empirical bond-order) potential (a version of the Abell-Tersoff-Brenner potentials \cite{abell1985prb,tersoff1988prb,brenner1990prb}) and a Lennard-Jones (LJ) potential for Mo-S systems. We call it the REBO-LJ potential from here on.  Stewart and Spearot \cite{stewart2013msmse} have slightly modified this potential and made an open source implementation \cite{rebolj_lammps} within LAMMPS. Here we present the version by Stewart and Spearot \cite{stewart2013msmse}.

For the REBO part, the total potential energy $U$ of the system can be written as a sum of the site potentials $U_i$:
\begin{equation}
U =   \sum_{i} U_i.
\end{equation}
The site potential takes the form:
\begin{equation}
U_i =  \frac{1}{2} \sum_{j \neq i} f_{\rm C}(r_{ij}) 
\left[ f_{\rm R}(r_{ij}) - b_{ij} f_{\rm A}(r_{ij}) \right],
\end{equation}
where $f_{\rm C}(r_{ij})$ is the Tersoff cutoff function \cite{tersoff1988prb}, $f_{\rm A}(r_{ij})$ and $f_{\rm R}(r_{ij})$ are respectively the attractive and the repulsive functions. The bond order function $b_{ij}$ is
\begin{equation}
b_{ij} =\left(1 + \zeta_{ij}\right)^{-1/2},
\end{equation}
where
\begin{equation}
\zeta_{ij} = \sum_{k\neq i, j} f_{\rm C}(r_{ik}) g(\cos\theta_{ijk}) + P(N_{i}).
\end{equation}
Here, $g$ is an analytical function of the bond angle $\theta_{ijk}$ formed by the $ij$ and $ik$ bonds, and $P$ is an analytical function of the coordination number $N_i$ defined as \cite{stewart2013msmse}
\begin{equation}
N_{i} = \sum_{j\neq i} f_{\rm C}(r_{ij}).
\end{equation}
Apart from the REBO part, a nonbonded LJ potential is also included to account for the van der Waals interactions. A cubic spline is constructed to smoothly reduce the LJ potential to zero at the inner cutoff distance of the REBO part. 

All material-specific parameters can be found in Ref. \onlinecite{stewart2013msmse}. The Lennard-Jones parameter $\epsilon$ for the S-S pair is set to $0.01386$ eV in Ref. \onlinecite{stewart2013msmse} and to $0.020$ eV in Refs. \onlinecite{liang2012prb}. We choose the value in Refs. \onlinecite{liang2012prb} because it is motivated by room temperature (300 K) applications whereas Ref. \onlinecite{stewart2013msmse} was motivated by zero temperature applications. 

\subsubsection{Original SW potential}

Before introducing the SW potentials for MoS$_2$, we review the original SW potential proposed in 1985 \cite{stillinger1985prb}. The total potential energy for the SW potential consists of a two-body part and a three-body part.
The site potential is
\begin{equation}
U_i = \frac{1}{2} \sum_{j\neq i} V_2(r_{ij}) + \frac{1}{2}\sum_{j\neq i}\sum_{k\neq i,j} h_{ijk},
\end{equation}
where
\begin{align}
V_2(r_{ij}) = &  A_{ij} \epsilon_{ij} \left[ B_{ij} \left( \frac{\sigma_{ij} }{ r_{ij} } \right)^{4}-1 \right] \nonumber \\
\times & \exp\left( \frac{1}{ r_{ij} / \sigma_{ij} - a_{ij}} \right)
\end{align}
and 
\begin{align}
h_{ijk}=&\epsilon_{ij}\lambda_{ijk}
\exp
\left[
\frac{\gamma_{ij}}{r_{ij}/\sigma_{ij}-a_{ij}} + \frac{\gamma_{ik}}{r_{ik}/\sigma_{ik}-a_{ik}}
\right] \nonumber \\
\times&\left(\cos \theta_{ijk} - \cos \theta_{0ijk} \right)^2.
\end{align}
Here, $A_{ij}$, $B_{ij}$, $\epsilon_{ij}$, $\sigma_{ij}$, $a_{ij}$, $\lambda_{ijk}$, $\gamma_{ij}$, and $\cos \theta_{0ijk}$ are material-specific parameters.
Parameters with two indices depend on a pair of atoms $i$ and $j$ (sometimes $i$ and $k$); parameters with three indices depend on a triplet $ijk$ of atoms $i$, $j$, and $k$, where $i$ is the central atom of the triplet. The parameter $\epsilon_{ij}$ is redundant and can be absorbed into $A_{ij}$ and $\lambda_{ijk}$. For each pair of atom types, there is a cutoff $\sigma_{ij}a_{ij}$ for the interactions.

\subsubsection{SW13 and SW13E potentials for molybdenum disulfide \label{section:sw13_sw13e_potentials}}

Jiang \textit{et al.} \cite{jiang2013jap} developed an SW potential in 2013 based on the standard SW potential and an extra requirement that there is no interaction in Mo-S-S triplets where the S-S distance is larger than the cutoff distance of $3.78$ \AA~ for the S-S pairs (such as the Mo-S$_1$-S$_5$ triplet in Fig. \ref{f_model}). The cutoff distance for the three-body part only extends to nearest neighbors (i.e., excluding Mo-Mo-Mo and S-S-S triplets), while that for the two-body part extends to next nearest neighbors (i.e., including Mo-S, Mo-Mo, and S-S pairs). However, the source code provided by Jiang \textit{et al.} (the file tagged with \verb"pair_sw.cpp" in the supplementary material of Ref. \onlinecite{jiang2013jap}) is incorrectly implemented such that all the three-body interactions are excluded. It is likely that this incorrect implementation has been used in some previous works on MoS$_2$. We call the potential as described in Ref. \onlinecite{jiang2013jap} (i.e, with the error in the source code fixed) the SW13 potential and that with the error in the source code unfixed the SW13E potential.  All the material-specific parameters can be found in Ref. \onlinecite{jiang2013jap}.

\subsubsection{SW16 potential for molybdenum disulfide}

Kandemir \textit{et al.} \cite{kandemir2016nt} developed another SW potential for MoS$_2$ in 2016, which we call the SW16 potential. This potential differs from the SW13 potential in that the Mo-S-S three-body interaction, for the case where the two S atoms are in the same sublayer (such as the Mo-S$_1$-S$_2$ triplet in Fig. \ref{f_model}), is taken to be different from the case where the two S atoms are in different sublayers (such as the Mo-S$_1$-S$_4$ triplet in Fig. \ref{f_model}). Similar to SW13, interactions in triplets such as Mo-S$_1$-S$_5$ in Fig. \ref{f_model} are excluded. All the material-specific parameters can be found in Ref. \onlinecite{kandemir2016nt}. 

\subsubsection{Time steps for integration}

By testing energy conservation in the NVE ensemble, we determined that a time step of $1.0$  fs is sufficiently small for the SW potentials but too large for the REBO-LJ potential, which requires a time step of $0.5$ fs to achieve good energy conservation. The need for a smaller time step for the REBO  potential originates from the Tersoff-like cutoff function adopted in the REBO-LJ potential, which is only continuous up to the first derivative.

\subsection{Methods for thermal conductivity calculations\label{section:method}}

There are multiple MD-based methods for heat transport calculations, including the equilibrium MD (EMD) method which is based on a Green-Kubo relation \cite{evans1990book} and the nonequilibrium MD (NEMD) method which is based directly on Fourier's law of heat conduction. When the simulation parameters (e.g. system size, simulation time, linear response) are properly chosen the two methods above are guaranteed to give consistent thermal conductivity results \cite{dong2018prb} in the diffusive regime. Additionally, there is a homogeneous nonequilibrium MD (HNEMD) method \cite{evans1982pla,evans1990book} which has been recently generalized \cite{fan2018submitted} such that it works for general many-body potentials, including the REBO and SW potentials considered in this work. When studying diffusive transport, the HNEMD and NEMD methods are the most and least computationally efficient, respectively \cite{xu2018submitted}. In this work, we mainly use the HNEMD method, but also employ the other two methods to crosscheck some results. We briefly review these methods below.

\subsubsection{The HNEMD method}

In this method, the system is driven out of equilibrium by an external force \cite{fan2018submitted}:
\begin{equation}
\vect{F}_{i}^{\rm ext}
= E_i \vect{F}_{\rm e} + \sum_{j \neq i} \left(\frac{\partial U_j}{\partial \vect{r}_{ji}} \otimes \vect{r}_{ij}\right) \cdot \vect{F}_{\rm e},
\end{equation}
where $\vect{r}_{ij}=\vect{r}_{j}-\vect{r}_{i}$, $\vect{r}_i$ being the position of particle $i$, $E_i$ is the total energy of atom $i$ and $\otimes$ denotes tensor product. When the parameter $\vect{F}_{\rm e}$ (of dimension inverse length) is small enough such that the system is in the linear response regime, a nonequilibrium heat current $\langle \vect{J} \rangle_{\rm ne}$, which is linear in $\vect{F}_{\rm e}$, will be induced. The linear relation between them can be expressed as 
\begin{equation}
\frac{\langle J^{\mu}(t)\rangle_{\rm ne}}{TV}
= \sum_{\nu} \kappa^{\mu\nu}(t) F_{\rm e}^{\nu},
\label{equation:kappa_munu}
\end{equation}
where $T$ is the system temperature, $V$ is the system volume and $\kappa^{\mu\nu}$ is the thermal conductivity tensor. For a many-body potential, the heat current $\vect{J}$ is given by \cite{fan2015prb}
\begin{equation}
\label{equation:J}
\vect{J} =\sum_i \vect{v}_i E_i + \sum_i \sum_{j \neq i} \vect{r}_{ij}
\left(
\frac{\partial U_j}{\partial \vect{r}_{ji}} \cdot \vect{v}_i
\right).
\end{equation}
Due to the hexagonal symmetry, the in-plane heat transport in MoS$_2$ is essentially isotropic. In this case, the in-plane thermal conductivity tensor reduces to a scalar $\kappa$ and can be expressed as 
\begin{equation}
\kappa(t)=\frac{\langle J(t)\rangle_{\rm ne}}{TVF_{\rm e}}.
\end{equation}
In practice, we redefine $\kappa(t)$ as the cumulative average of the above quantity \cite{fan2018submitted}:
\begin{equation}
\label{equation:running_average}
\kappa(t) = \frac{1}{t}\int_0^t \frac{\langle J(s)\rangle_{\rm ne}}{TVF_{\rm e}}ds,
\end{equation}
and check how $\kappa(t)$ convergences.
More theoretical and technical details for the HNEMD method can be found from Ref. \onlinecite{fan2018submitted}.

The simulation protocol in the HNEMD method is as follows. First, we equilibrate the system for $1$ ns in the NPT ensemble with a target temperature of $300$ K and a target in-plane pressure of zero. Second, we make a production run in the NVT ensemble, measuring and outputting the average heat current for every $1000$ steps.
As we will see, the thermal conductivities we calculate using the SW potentials are much larger than what we calculate using the REBO-LJ potential.
Therefore, the production time for the SW potentials also needs to be much larger. We use a production time of $2$ ns for the REBO-LJ potential and a production time of $15$ ns for the SW potentials. Accordingly, the driving force parameter $F_{\rm e}$ needs to be smaller for the SW potentials. With some tests, we determined the following appropriate parameters: $F_{\rm e}=0.2$ $\mu$m$^{-1}$ for the REBO-LJ potential and $0.05$ $\mu$m$^{-1}$ for the SW potentials. Last, as this method has small finite-size effects (because there is no boundary scattering), a relatively small simulation cell can be used. We tested two cell sizes, $16 \times 16$ nm$^2$ ($N=9000$ atoms) and $32 \times 32$ nm$^2$ ($N=36000$ atoms), with periodic boundary conditions in the $xy$ plane and obtained consistent results. This is also the case for the EMD method introduced below. We performed $10$ independent runs and calculated the error bounds in terms of the standard error (standard deviation divided by the square root of the number of independent runs).

\subsubsection{The EMD method}

In the EMD method, one first calculates the ensemble (time) average $\langle \cdots \rangle$ of the heat current autocorrelation function
$\langle J(t)J(0)\rangle$ and then performs a numerical integration to get the (running) thermal conductivity $\kappa(t)$ according to the following Green-Kubo relation \cite{evans1990book,tuckerman2010book}:
\begin{equation}
\label{equation:gk}
\kappa(t)=\frac{1}{k_{\rm B}T^2V}\int_0^t dt' \langle J(t')J(0)\rangle,
\end{equation}
where $k_{\rm B}T$ is the thermal energy  and $V$ is the volume of the system. In practice, one needs to check the time convergence of the running thermal conductivity. This is called the EMD method because the heat current here is sampled in equilibrium state (in the NVE ensemble). 

We only used the EMD method to crosscheck some results obtained by using the SW potentials. The simulation protocol is as follows. First, we equilibrate the system for $1$ ns in the NPT ensemble with a target temperature of $300$ K and a target in-plane pressure of zero. Second, we make a production run of $50$ ns in the NVE ensemble, sampling the instant heat current every $10$ steps. Third, we calculate the heat current autocorrelation function using the saved heat current data and then calculate the running thermal conductivity according to the Green-Kubo relation Eq. (\ref{equation:gk}). We performed $50$ to $100$ independent runs and calculated the error bounds in terms of the standard error. 

\subsubsection{The NEMD method}

The NEMD method can be used to calculate the thermal conductivity $\kappa(L)$ of systems with finite length $L$ according to Fourier's law:
\begin{equation}
\kappa(L)=\frac{Q}{|\nabla T|},
\label{equation:kappa_NEMD}
\end{equation}
where $Q$ is an externally generated heat flux and $\nabla T$ is the resulting temperature gradient in steady state. There are many flavors of the NEMD method and we chose the following setup: we fix the two ends of the system in the transport direction and generate the heat flux by maintaining the temperatures in the local atomic groups close to the left and the right ends at $330$ K (heat source) and $270$ K (heat sink), respectively, using the Nos\'e-Hoover chain method \cite{tuckerman2010book}. The heat flux is calculated based on energy conservation between the system and the baths.

Again, here we only used the NEMD method to crosscheck some results obtained using the SW16 potential. The simulation protocol is as follows: First, we equilibrate the system for $1$ ns in the NPT ensemble with a target temperature of $300$ K and a target in-plane pressure of zero. Second, we make a production run of $10$ ns with local thermal baths, sampling the local temperatures and the accumulated energy exchanged between the system and the thermal baths. Third, we use the data within the last $6$ ns of the production stage (where we checked that steady state has been achieved) to determine the temperature gradient $\nabla T$, the energy exchange rate $dE/dt$, and the heat flux $Q=dE/dt/S$, where $S$ is the cross-sectional area. The thermal conductivity is then calculated according to Eq. (\ref{equation:kappa_NEMD}). We keep the width of the system at $10$ nm and vary the length from $200$ nm to $1000$ nm. Periodic boundary conditions are applied to the width direction. For each length, we performed three independent runs and calculated the error bounds in terms of the standard error. 

\section{Results and discussion\label{section:results}}

\subsection{HNEMD results for single-layer molybdenum disulfide \label{section:hnemd}}

\begin{figure*}[htb]
\begin{center}
\includegraphics[width=1.5\columnwidth]{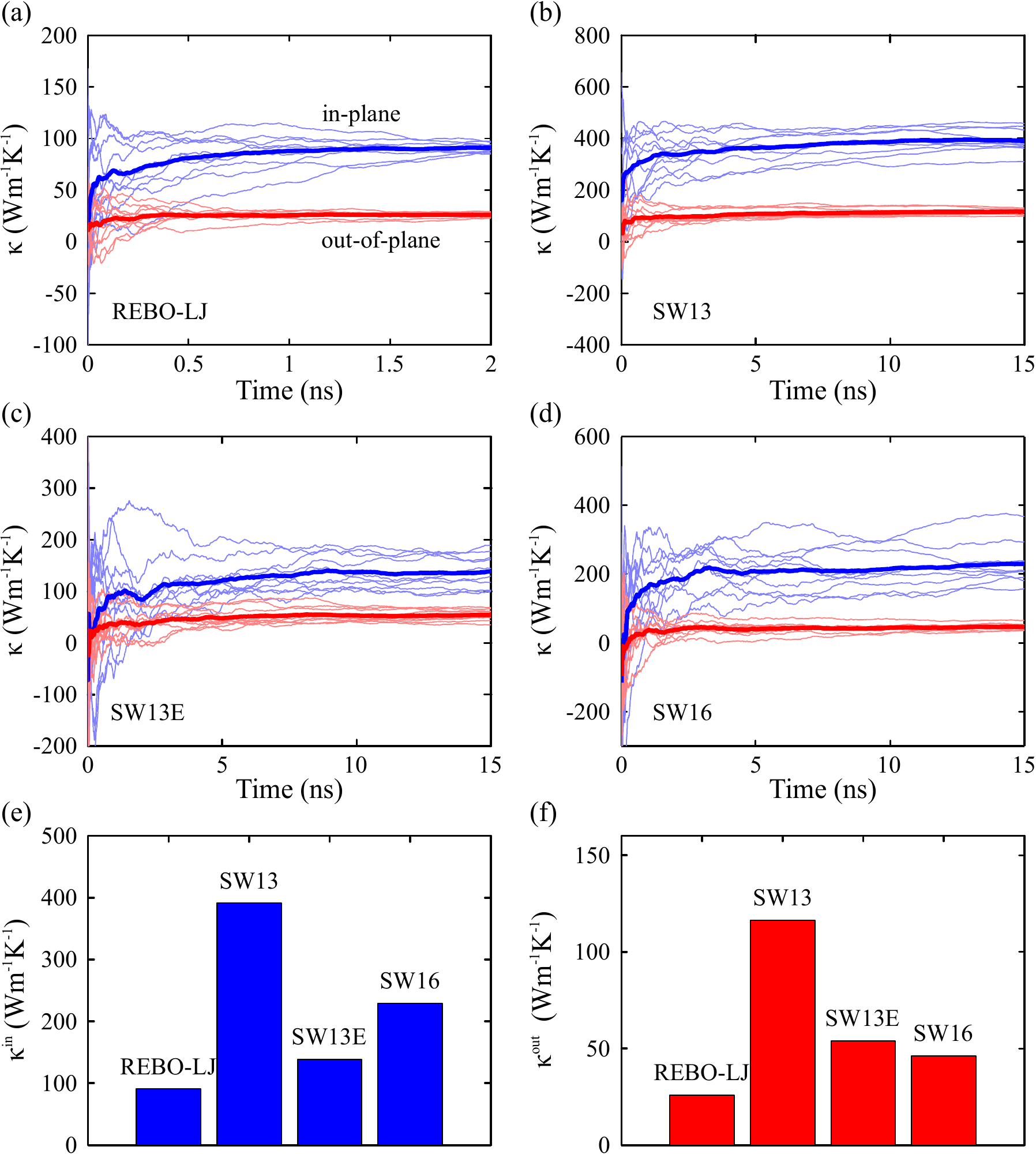}
\caption{(a) to (d) Cumulative averages of the thermal conductivity [Eq. (\ref{equation:running_average})] as a function of time from the HNEMD simulations for single-layer MoS$_2$ at $300$ K and zero pressure using the various potentials. The thin lines in each subplot represent ten independent runs and the thick line is the average. (e) and (f) Converged thermal conductivity values for the different potentials.}
\label{f_kappa_hnemd}
\end{center}
\end{figure*}

The accumulative averages of the in-plane thermal conductivity in suspended single-layer MoS$_2$ calculated using the HNEMD method [cf. Eq. (\ref{equation:running_average})] with the various potentials are shown in Fig. \ref{f_kappa_hnemd}. Here, we have decomposed the thermal conductivity into an in-plane component and an out-of-plane component, $\kappa=\kappa^{\rm in}+\kappa^{\rm out}$, which corresponds to the following decomposition of the heat current \cite{fan2017prb}:
\begin{equation}
\vect{J} = \vect{J}^{\rm in} + \vect{J}^{\rm out};
\end{equation}
\begin{equation}
\vect{J}^{\rm in} 
= \sum_i  \sum_{j\neq i} \vect{r}_{ij} 
\left(
  \frac{\partial U_j}{\partial x_{ji}} v^{x}_{i}+
  \frac{\partial U_j}{\partial y_{ji}} v^{y}_{i}
\right);
\end{equation}
\begin{equation}
\vect{J}^{\rm out} 
= \sum_i \sum_{j\neq i} \vect{r}_{ij} 
  \frac{\partial U_j}{\partial z_{ji}} v^{z}_{i}.
\end{equation}
We note that the out-of-plane component $\kappa^{\rm out}$ is not the thermal conductivity in the cross-plane direction of the MoS$_2$ layer, but the in-plane thermal conductivity contribution from out-of-plane phonon modes. It is interesting to note that, for all the potentials, the thermal conductivity is dominated by the in-plane component. This is similar to black phosphorous \cite{xu2018submitted}, but opposite to graphene \cite{lindsay2010prb,fan2017prb} and h-BN \cite{dong2018submitted}. In planar 2D materials such as graphene and h-BN, the out-of-plane modes correspond exactly to the flexural modes and there is a symmetry selection rule \cite{lindsay2010prb} that excludes three-phonon scattering processes involving an odd number of flexural phonons. This leads to much larger relaxation times for the flexural phonon modes which are consequently the major heat carriers in these materials. Because MoS$_2$ is not a strictly planar (one-atomic-thick) crystal, the symmetry selection rule does not apply, which leads to  relatively stronger phonon-phonon scattering rates and a relatively smaller thermal conductivity contribution from the out-of-plane modes \cite{gu2018rmp}.

We report the converged thermal conductivity values for the various potentials from the HNEMD method in Table \ref{t_hnemd}. The large thermal conductivity values computed using the SW13 and SW16 potentials are clearly unphysical as compared to experimental data, while that computed using the REBO-LJ potential is very reasonable. Note that our HNEMD predictions using the SW potentials differ significantly from those from previous works \cite{ding2015nt,jin2015sr,kandemir2016nt,hong2016jpcc}; see Table \ref{t_SW}. We give the detailed comparisons next.

\begin{table}[ht]
\caption{Thermal conductivity values (in units of W m$^{-1}$ K$^{-1}$) for single-layer MoS$_2$ at $300$ K and zero pressure from the HNEMD simulations. }
\begin{center}
\begin{tabular}{llll}
\hline
\hline
Potential & $\kappa^{\rm in}$ & $\kappa^{\rm out}$ & $\kappa$  \\
\hline
REBO-LJ & $91   \pm 2$   & $26   \pm 1$  & $117  \pm 3$  \\  
SW13    & $391  \pm 14$  & $116  \pm 3$  & $507  \pm 17$ \\  
SW13E   & $139  \pm 11$  & $54   \pm 3$  & $193  \pm 14$ \\   
SW16    & $229  \pm 19$  & $46   \pm 3$  & $275  \pm 22$ \\  
\hline
\end{tabular}
\end{center}
\label{t_hnemd}
\end{table}

\subsection{Comparison with previous MD results
\label{section:comparison}}

\begin{table}[ht]
\caption{Thermal conductivity values (in units of W m$^{-1}$ K$^{-1}$) for single-layer MoS$_2$ at $300$ K and zero pressure as calculated in this work and from previous works using the SW potentials. }
\begin{center}
\begin{tabular}{llll}
\hline
\hline
Reference & Potential & Method & $\kappa $  \\
\hline
\cite{ding2015nt}    & SW13/SW13E    & NEMD  & $19.76$  \\  
\cite{jin2015sr}     & SW13/SW13E    & EMD  & $116.8$ \\  
Here & SW13          & HNEMD         & $507\pm 17$ \\
Here & SW13          & EMD (GPUMD)   & $531\pm53$\\
Here & SW13          & EMD (LAMMPS)  & Diverged\\
Here & SW13E         & HNEMD         & $193\pm 14$ \\
Here & SW13E         & EMD (GPUMD)   & $209\pm22$\\
Here & SW13E         & EMD (LAMMPS)  & $208\pm13$\\
\cite{kandemir2016nt}& SW16  & Einstein relation  & $95  \pm 5$ \\  
\cite{hong2016jpcc}  & SW16   & EMD   & $108.74 \pm 6.68$ \\  
\cite{hong2016jpcc}  & SW16   & NEMD  & $110.30 \pm 2.07$ \\  
Here                 & SW16   & HNEMD & $275 \pm 22$\\
Here                 & SW16   & EMD (GPUMD)  & $280 \pm 32$\\
Here                 & SW16   & EMD (LAMMPS)  & $231 \pm 16$\\
Here                 & SW16   & NEMD  & $262 \pm 28$\\
\hline
\end{tabular}
\end{center}
\label{t_SW}
\end{table}

\subsubsection{Comparison with previous MD results using the SW13/SW13E potential}

\begin{figure*}[htb]
\begin{center}
\includegraphics[width=1.4\columnwidth]{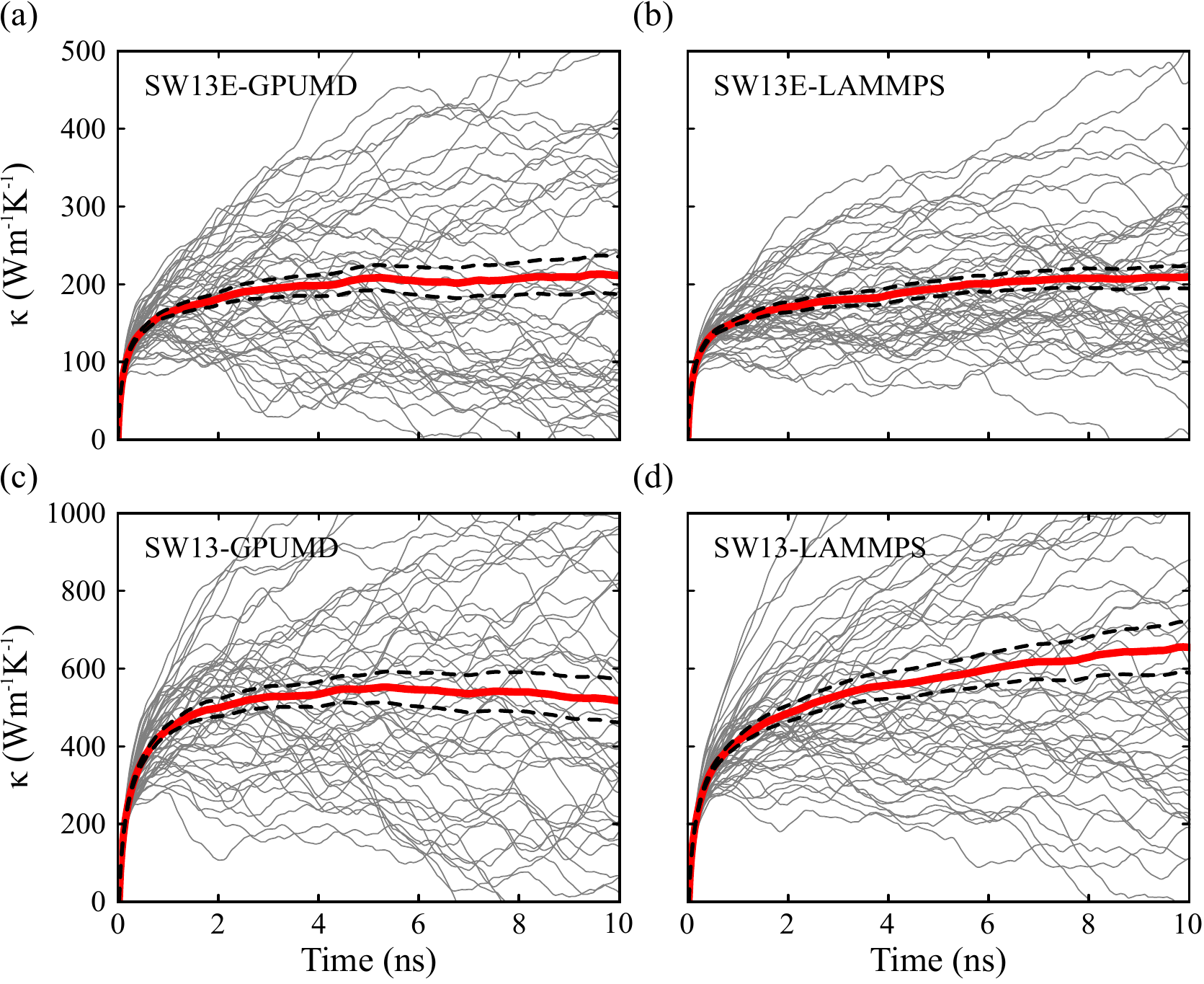}
\caption{EMD results for single-layer MoS$_2$ at 300 K and zero pressure. (a), (b) show results using the SW13E potential for GPUMD and LAMMPS respectively. (c), (d) show results using the SW13 potential for GPUMD and LAMMPS respectively. The thin gray lines represent the results of independent simulations with different initial velocities ($50$ for each plot), the red thick solid lines indicate the average of all independent simulations, and the black thick dotted lines show the error bounds.}
\label{f_kappa_emd_sw13e-sw13}
\end{center}
\end{figure*}

Our HNEMD values for the SW13 and SW13E potentials are $507 \pm 17$ and $193 \pm 14$ W m$^{-1}$ K$^{-1}$, respectively. In comparison, Ding \textit{et al.} \cite{ding2015nt} obtained a value of $19.76$ W m$^{-1}$ K$^{-1}$ using the NEMD method and Jin \textit{et al.}  \cite{jin2015sr} obtained a value of $116.8$ W m$^{-1}$ K$^{-1}$ using the EMD method. Unfortunately, it is not clear whether they have used the SW13 or the SW13E potential. In any case, the results by Ding \textit{et al.} \cite{ding2015nt} can be understood by noticing that they have used very short system lengths in their NEMD simulations, which has been demonstrated \cite{sellan2010prb,dong2018prb} to be inadequate for making a reliable extrapolation to the limit of infinite length.

The EMD results by Jin \textit{et al.}  \cite{jin2015sr} differ from our HNEMD results obtained with both the SW13 and the SW13E potentials. To resolve this discrepancy, we performed EMD simulations using both the GPUMD code and the LAMMPS code with these two potentials. For the SW13E potential, we find that the thermal conductivity converges to similar values at $209 \pm 22$ W m$^{-1}$ K$^{-1}$ and $208 \pm 13$ W m$^{-1}$ K$^{-1}$ for GPUMD and LAMMPS respectively (Figs. \ref{f_kappa_emd_sw13e-sw13}(a) and (b) respectively). As discussed in Sec. \ref{section:sw13_sw13e_potentials}, the SW13E potential excludes all of the three-body interactions. Under this circumstance, the heat current calculations for GPUMD and LAMMPS are the same and an identical thermal conductivity between the codes is expected. Our EMD values are also consistent with our HNEMD value for the SW13E potential. 

With the SW13 potential, three-body interactions are included and the heat current calculations in GPUMD and LAMMPS are different \cite{fan2015prb}. We find these differences to be substantial as the thermal conductivity is calculated to be $531\pm53$ W m$^{-1}$ K$^{-1}$ using GPUMD and the (incorrect) LAMMPS simulations \textit{do not converge} up to a correlation time of $10$ ns (Figs. \ref{f_kappa_emd_sw13e-sw13}(c) and (d) respectively). Among the four combinations of software packages and the SW13E/SW13 potentials, we do not obtain results that are consistent with those in Ref. \onlinecite{jin2015sr}, leaving the discrepancy unresolved. Again, our EMD value using GPUMD is consistent with our HNEMD value for the SW13 potential. We also note that the HNEMD method is more than an order of magnitude more efficient than the EMD method, as has been demonstrated for many other systems \cite{fan2018submitted,xu2018submitted,dong2018submitted}.

\subsubsection{Comparison with previous MD results using the SW16 potential}

Our HNEMD thermal conductivity for the SW16 potential, $275 \pm 22$ W m$^{-1}$ K$^{-1}$, is much larger than those in Refs. \onlinecite{kandemir2016nt,hong2016jpcc} [about $100$ W m$^{-1}$ K$^{-1}$; cf. Table \ref{t_SW}]. They have used the EMD, the NEMD, and the Einstein-relation \cite{kinaci2012jcp}  methods and it seems that their results agree with each other very well. To understand the discrepancies, we performed EMD and NEMD simulations using this potential. 

\begin{figure*}[htb]
\begin{center}
\includegraphics[width=1.5\columnwidth]{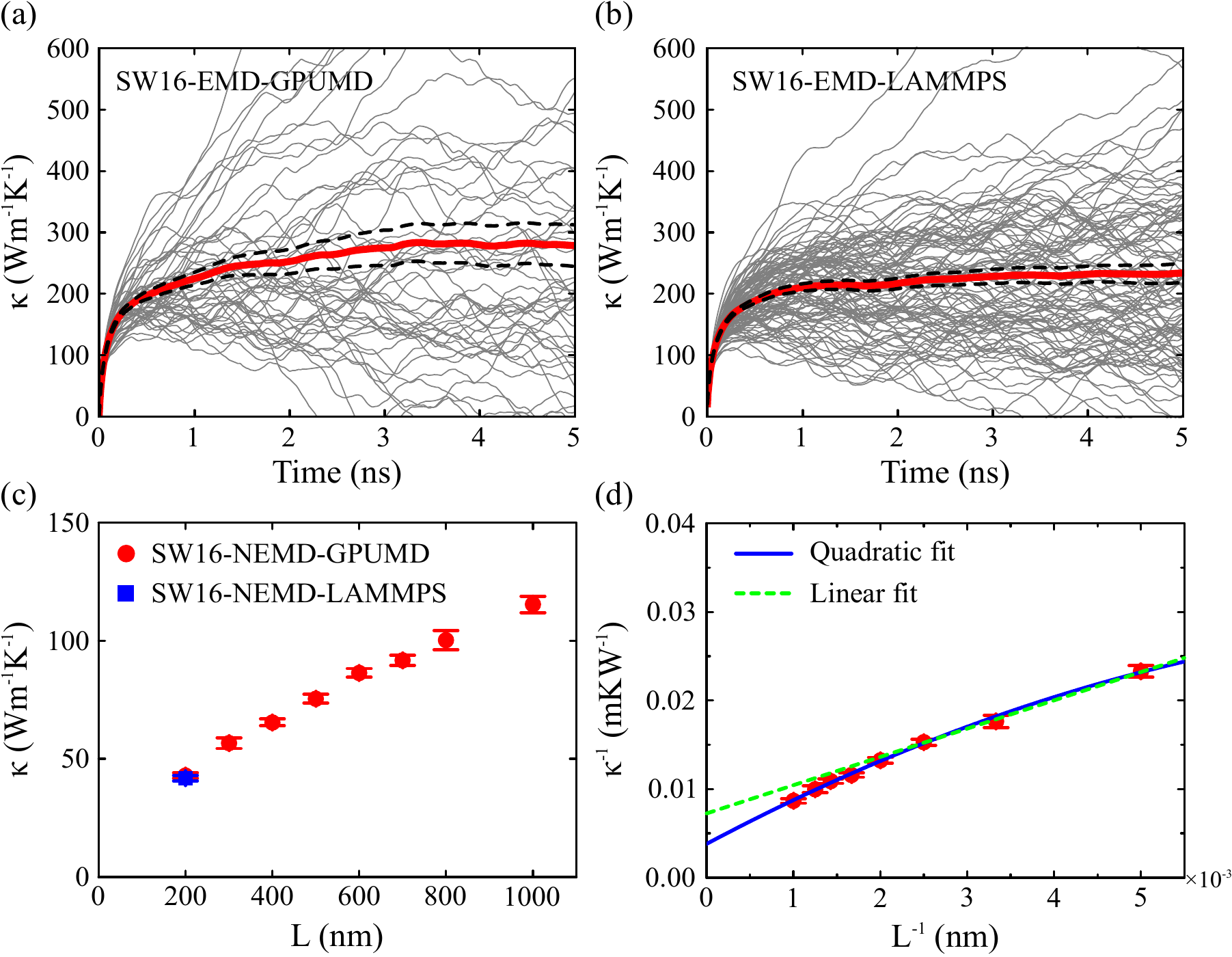}
\caption{EMD and NEMD results for single-layer MoS$_2$ at 300 K and zero pressure using the SW16 potential. Running thermal conductivity as a function of correlation time from the EMD simulations using GPUMD (a) and LAMMPS (b). In (a) and (b), the thin lines represent the results from independent runs and the thick line represents their average. The dashed lines represent error bounds. (c) $\kappa(L)$ from the NEMD simulations. (d) $1/\kappa(L)$ as a function of $1/L$, where the solid line represents a quadratic fit [Eq. (\ref{equation:kappa_fit})] to all the data points and the dashed line represents a linear fit [Eq. (\ref{equation:kappa_fit}) with $\alpha=0$] to the data with $L=200$ nm to $400$ nm.}
\label{f_sw16}
\end{center}
\end{figure*}

Figures \ref{f_sw16}(a) and \ref{f_sw16}(b) show the EMD results obtained by using the GPUMD code and the LAMMPS code, respectively. Similar to the case of the SW13 potential, the convergence of $\kappa(t)$ with respect to $t$ is very slow, indicating the inadequate crystal anharmonicity represented by this potential. The converged thermal conductivity using GPUMD is $\kappa=280\pm 32$ W m$^{-1}$ K$^{-1}$, which is consistent with our HNEMD prediction. The converged thermal conductivity using LAMMPS is $\kappa=231\pm 16$ W m$^{-1}$ K$^{-1}$, which is smaller than the GPUMD value, but much larger than the EMD value by Hong \textit{et al.} \cite{hong2016jpcc}. It has been found that the heat current formula in LAMMPS leads to significantly underestimated thermal conductivity for various two-dimensional materials described by many-body potentials, including graphene \cite{fan2015prb}, h-BN \cite{dong2018submitted}, and black phosphorous \cite{xu2018submitted}. Here, the difference between our GPUMD and LAMMPS results is small, which might be related to the fact that the SW potentials for MoS$_2$ contain a large portion of the two-body component, as explained in Sec. \ref{section:potential}. 

\begin{table}[hbt]
\caption{Thermal conductivity values (in units of W m$^{-1}$ K$^{-1}$) calculated with different packages for systems with different lengths  $L$ (in units of nm) in the NEMD simulations using the SW16 potential. The number of atoms in each system is denoted as $N$.}
\begin{center}
\begin{tabular}{c c c c}
\hline
\hline
$L$ &   $N$     &   $\kappa$            &   package \\
\hline
200	&	72162	&	$42   \pm   0.6$    &	LAMMPS	\\
200	&	72162	&	$43   \pm   1.2$	&	GPUMD	\\
300	&	108186	&	$57   \pm 	2.2$	&	GPUMD	\\
400	&	144324	&	$65   \pm 	1.5$	&	GPUMD	\\
500	&	180384	&	$75   \pm 	1.8$	&	GPUMD	\\
600	&	216486	&	$86   \pm 	1.9$	&	GPUMD	\\
700	&	252510	&	$92   \pm	2.1$	&	GPUMD	\\
800 &	288648	&	$100  \pm   4.1$	&	GPUMD	\\
1000&	360810	&	$115  \pm   3.4$	&	GPUMD	\\
\hline
\end{tabular}
\end{center}
\label{t_nemd_sw16}
\end{table}

Because we failed to reproduce the EMD results by Hong \textit{et al.} \cite{hong2016jpcc} using the LAMMPS code, we further tried the NEMD method. 
Using this method and the GPUMD code, we calculated the thermal conductivity $\kappa(L)$ of MoS$_2$ with the length $L$ varying from $200$ nm to $1000$ nm. The data are listed in Table \ref{t_nemd_sw16} and  shown in Fig. \ref{f_sw16}(c). For the case of $L=200$ nm, we have used both GPUMD and LAMMPS and got identical results. To see how $\kappa(L)$ converges with increasing $L$, we plot $1/\kappa(L)$ against $1/L$ in Fig. \ref{f_sw16}(d). Usually, it is assumed \cite{Schelling2002prb} that $1/\kappa(L)$ is linear in $1/L$, but in most cases $1/\kappa(L)$ is a nonlinear function of $1/L$ \cite{dong2018prb,sellan2010prb} due to the frequency dependence of the phonon mean free path $\lambda(\omega)$. Our NEMD data can be well fitted by a quadratic function [the solid line in Fig. \ref{f_sw16}(d)]:
\begin{equation}
\frac{1}{\kappa(L)}=\frac{1}{\kappa_0}
\left( 1 + \frac{\lambda}{L} +\frac{\alpha}{L^2}\right),
\label{equation:kappa_fit}
\end{equation}
where $\lambda \approx 1370$ nm is the effective (average) phonon mean free path and $\alpha $ is a (negative) parameter characterizing the nonlinearity caused by the frequency dependence of the phonon mean free path. The thermal conductivity in the infinite-length limit $\kappa_0$ is fitted to be $262 \pm 28$ W m$^{-1}$ K$^{-1}$. This is close to but slightly smaller than our HNEMD and EMD predictions. This is because the maximum length ($1000$ nm) we have considered in our NEMD simulations is still shorter than $\lambda$ and is not long enough \cite{sellan2010prb,dong2018prb} to fully capture the nonlinearity between $1/\kappa(L)$ and $1/L$. We think this could explain why Hong \textit{et al.} \cite{hong2016jpcc} obtained a much smaller value ($110.30 \pm 2.07$ W m$^{-1}$ K$^{-1}$) using NEMD simulations. Their NEMD data actually exhibit nonlinear dependence between $1/\kappa(L)$ and $1/L$, but they still have used a linear fit to their data up to a length of $400$ nm. Actually, by fitting our data up to the same length using a linear function (cf. the dashed line in Fig. \ref{f_sw16}(d)), we get an extrapolated value of $\kappa_0 = 137 \pm 7$, which is close to the value reported by Hong \textit{et al.} \cite{hong2016jpcc}.  

The developers of the SW16 potential \cite{kandemir2016nt} have calculated the thermal conductivity of single-layer MoS$_2$ using the so-called Einstein-relation method \cite{kinaci2012jcp}. They obtained a value of $95 \pm 5$ W m$^{-1}$ K$^{-1}$, which is significantly smaller than ours. No details regarding the method and the time convergence in their calculations were presented, and we do not know the origin of the discrepancy. However, we note that the Einstein-relation method consistently underestimated the thermal conductivity of some other materials: it gives a value of $160.5 \pm 10.0$ W m$^{-1}$ K$^{-1}$ for silicon at 300 K (using the Tersoff potential \cite{tersoff1989prb}) against a value of $250 \pm 10$ W m$^{-1}$ K$^{-1}$ using the standard EMD method \cite{dong2018prb}; it gives a value of $400$ W m$^{-1}$ K$^{-1}$ for single-layer h-BN at $300$ K (using the Tersoff potential \cite{sevik2011prb}) against a value of $670 \pm 30$ W m$^{-1}$ K$^{-1}$ using both the EMD and the HNEMD method \cite{dong2018submitted}.

\subsection{Comparison among the empirical potentials and with experiments\label{section:experiments}}

In this subsection, we give detailed comparisons between our simulation results and available experimental data, as summarized in Table \ref{table:kappa}. Experimentally measured in-plane thermal conductivities from various sources are typically smaller than $100$ W m$^{-1}$ K$^{-1}$. Only the prediction by the REBO-LJ potential is close to these; the SW potentials significantly overestimate the thermal conductivity due to the underestimated phonon anharmonicity, as evidenced from the extraordinarily slow convergence of the running thermal conductivity (see Fig. \ref{f_kappa_emd_sw13e-sw13}(c) and Fig. \ref{f_sw16}(a)). 
In contrast, both the time scale and the thermal conductivity value from the REBO-LJ potential are very reasonable (see Fig. \ref{f_kappa_hnemd}(a)).

\begin{table}[htb]
\caption{In-plane thermal conductivity values for suspended and bulk MoS$_2$ at $300$ K from experiments and our predictions using the HNEMD method with the REBO-LJ potential. Isotope scattering is considered in the HNEMD calculations. }
\begin{center}
\begin{tabular}{ l l l l }
\hline
\hline
Ref.                       & Sample or method    & Layers & $\kappa$ (W m$^{-1}$ K$^{-1}$) \\
\hline
\cite{yarali2017afm}       & CVD                  &1      & $30 \pm 3.3$; $35.5 \pm 3$ \\
\cite{bae2017nanoscale}    & CVD                  &1      & $13.3 \pm 1.4$\\
\cite{bae2017nanoscale}    & CVD                  &2      & $15.6 \pm 1.5$\\
\cite{sahoo2013jpcc}       & CVD                  &11     & $52$ \\
\cite{bae2017nanoscale}    & CVD                  &12     & $43.4 \pm 9.1$\\
\cite{yan2014acsnano}      & Exfoliated           &1      & $34.5 \pm 4$\\
\cite{zhang2015acsami}     & Exfoliated           &1      & $84 \pm 17$\\
\cite{zhang2015acsami}     & Exfoliated           &2      & $77 \pm 25$\\
\cite{aiyiti2018nanoscale} & Exfoliated           &4      & $34 \pm 5$; $31 \pm 4$ \\
\cite{jo2014apl}           & Exfoliated           &4      & $44-50$ \\
\cite{jo2014apl}           & Exfoliated           &7      & $48-52$ \\
\cite{liu2014jap}          & Natural crystal      &bulk   & $85-110$ \\
\cite{zhu2016nc}          & Natural crystal      &bulk   & $105$ \\
\cite{jiang2017am}         & Natural crystal      &bulk   & $80 \pm 17$ \\
Here & HNEMD (REBO-LJ)  & 1    & $110\pm 4$ \\
Here & HNEMD (REBO-LJ)  & 2    & $92\pm 4$  \\
Here & HNEMD (REBO-LJ)  & 3    & $81\pm 3$  \\
Here & HNEMD (REBO-LJ)  & 4    & $78\pm 3$  \\
Here & HNEMD (REBO-LJ)  & 5    & $80\pm 3$  \\
Here & HNEMD (REBO-LJ)  & bulk & $83\pm 3$  \\
\hline
\end{tabular}
\end{center}
\label{table:kappa}
\end{table}

A proper description of thermal transport without an adequate description of phonon dispersion is impossible. To this end, we also calculated the phonon dispersions of MoS$_2$ as described by the empirical potentials using the finite displacement method as implemented in the PHONOPY \cite{togo2015sm} package and compared with experiment data \cite{tornatzky2018arxiv} determined by inelastic X-ray scattering. The required harmonic force constants are calculated by using a $8 \times 5$ rectangular supercell (240 atoms), which is large enough to take care of the long-range LJ potential in the REBO-LJ potential. The SW potentials do not need a supercell as large as this but we have used this supercell uniformly for all the potentials. Before generating the displacements, the supercell has been optimized at zero temperature for each potential. Phonon dispersion curves in the Brillouin zone corresponding to the primitive cell were obtained by unfolding \cite{UNFOLDING} those corresponding to the supercell.

\begin{figure*}[htb]
\begin{center}
\includegraphics[width=1.5\columnwidth]{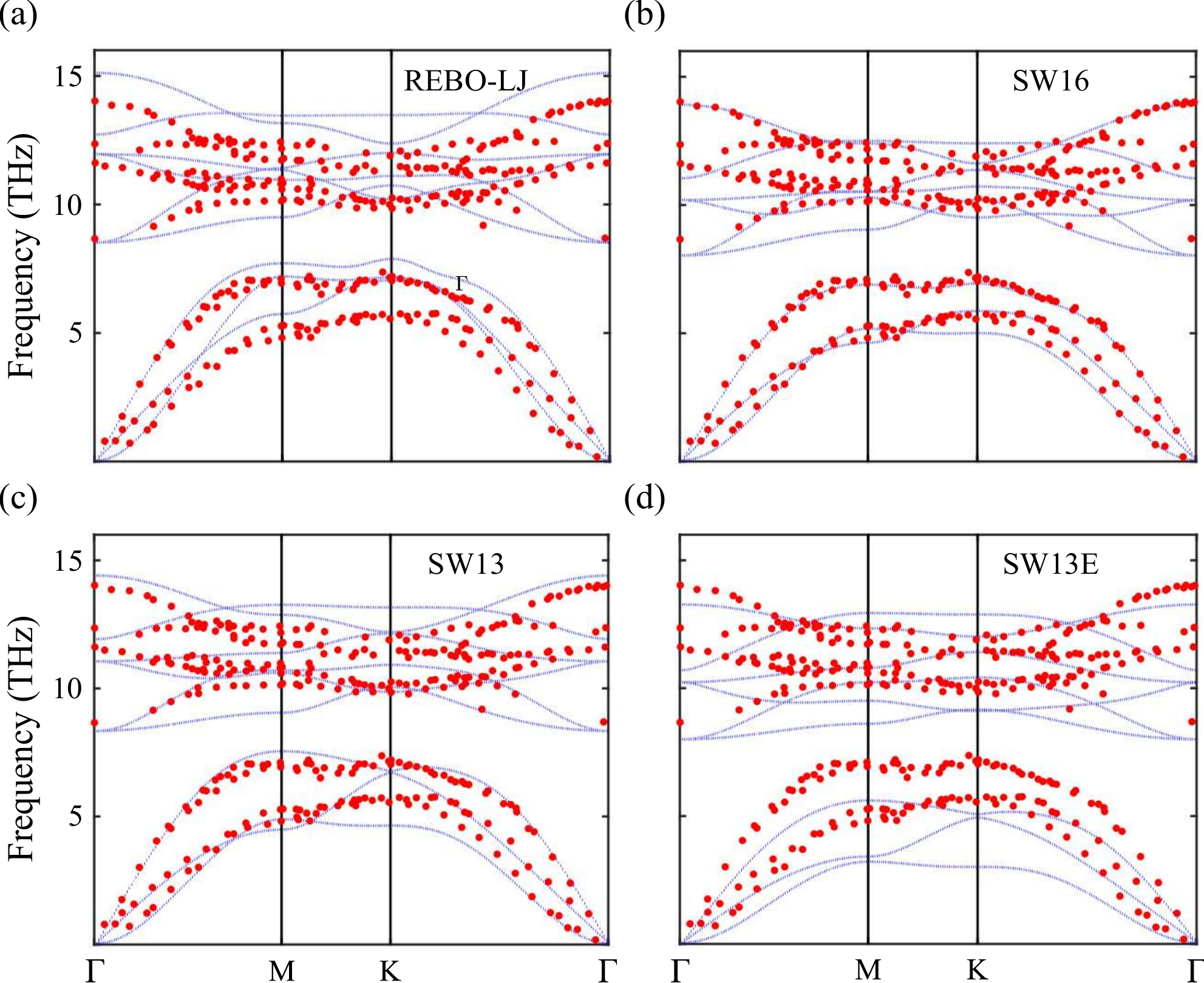}
\caption{Phonon dispersion curves for single-layer MoS$_2$ along the $\Gamma$-M-K-$\Gamma$ high symmetry path calculated using the (a) REBO-LJ potential, (b) SW16 potential, (c) SW13 potential, and (d) SW13E potential.
Red dots are experiment data for bulk MoS$_2$ \cite{tornatzky2018arxiv}.}
\label{phon}
\end{center}
\end{figure*}

The calculated phonon dispersions are shown in Fig. \ref{phon}. By comparing with the experimental data \cite{tornatzky2018arxiv}, we see that all the three potentials (REBO-LJ, SW13, and SW16) describe the low-frequency acoustic phonons fairly well. The SW13E potential does not lead to reasonable description of the phonon dispersion, which is expected as it is an incorrect implementation of the SW13 potential.

From the comparisons above between the various potentials, in terms of thermal conductivity and phonon dispersion of single-layer MoS$_2$, we conclude that the REBO-LJ potential stands out. The REBO-LJ potential has two other advantages: First, it is more transferable because it was fitted by considering various Mo-S structures with diverse coordination numbers. An evidence for its high transferability comes from a very satisfactory reproduction \cite{Ghorbani-Asl2017} of the formation energies of point defects in MoS$_2$, as compared to those obtained using DFT calculations \cite{KomsaPRB2015}. In contrast, the SW potentials \cite{jiang2013jap,kandemir2016nt} were fitted by considering only some equilibrium properties such as bond lengths and elastic constants. The implementations of these SW potentials also involve an \textit{ad hoc} modification to the LAMMPS source code, which is problematic when the structure is away from the equilibrium MoS$_2$ structure; the intention of this modification was to exclude the interactions in triplets such as Mo-S$_1$-S$_5$ in Fig. \ref{f_model}, but it will lead to excluding the S-Mo-Mo triplet interactions when the Mo-Mo distance is larger than the S-S cutoff distance and smaller than the Mo-Mo cutoff distance, which is unreasonable. Second, there is an intrinsic van der Waals part in the REBO-LJ potential, which is important for describing multi-layer systems. This part was not included in the SW potentials. From here on, we only use the REBO-LJ potential and the efficient HNEMD method, focusing on comparisons with experiments \cite{muratore2013apl,liu2014jap,zhu2016nc,jiang2017am,sahoo2013jpcc,jo2014apl,yan2014acsnano,zhang2015acsami,bae2017nanoscale,yarali2017afm,aiyiti2018nanoscale} and results from BTE approach combined with DFT calculations \cite{li2013apl,gu2016jap}.

\begin{figure*}[htb]
\begin{center}
\includegraphics[width=1.5\columnwidth]{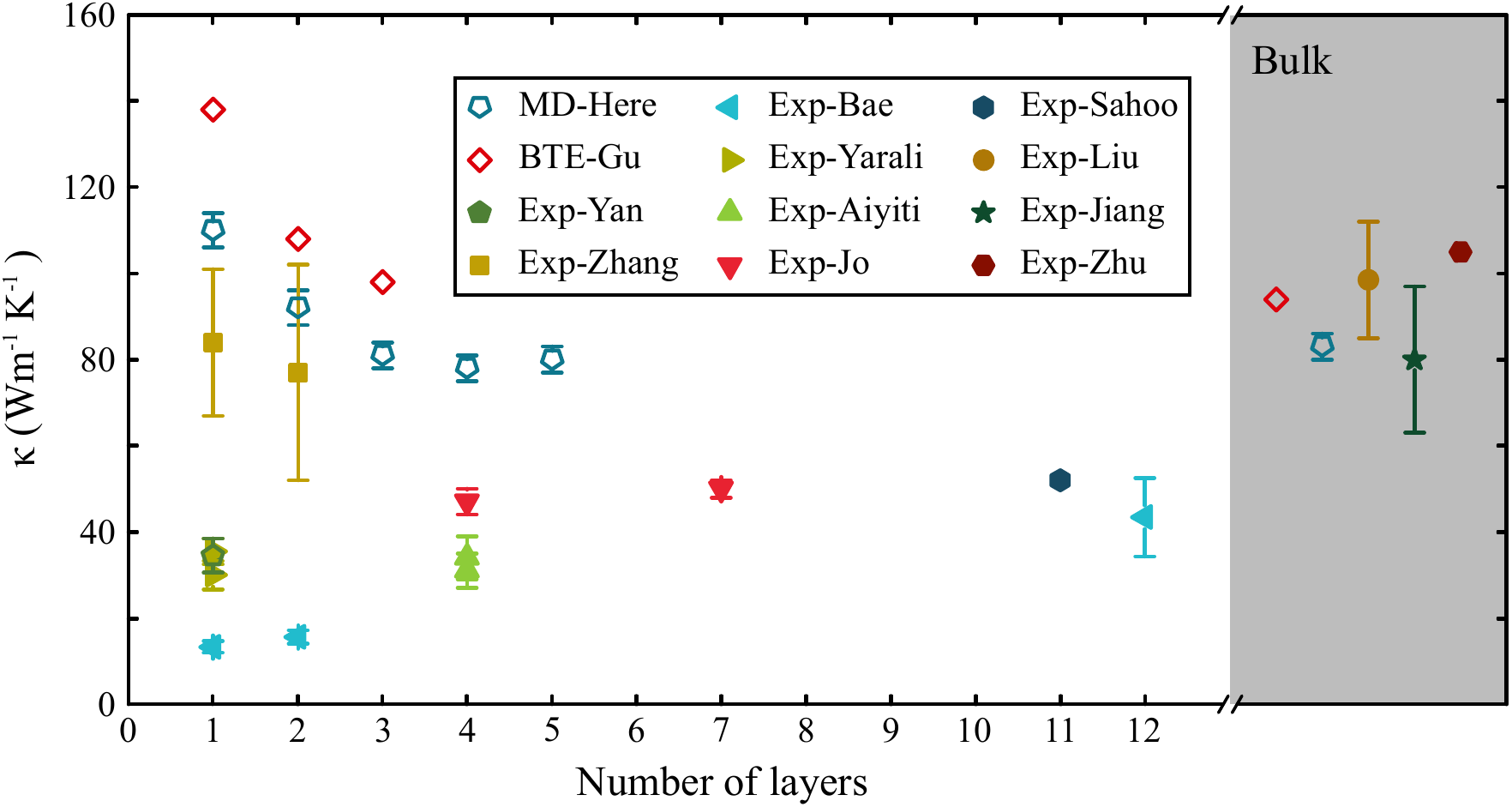}
\caption{Thermal conductivity as a function of the number of layers for MoS$_2$ at $300$ K and zero pressure. Sources of reference data:  
Liu \cite{liu2014jap};
Zhu \cite{zhu2016nc}; 
Jiang \cite{jiang2017am}; 
Sahoo \cite{sahoo2013jpcc}; 
Jo \cite{jo2014apl};
Yan \cite{yan2014acsnano}; 
Zhang \cite{zhang2015acsami}; 
Bae \cite{bae2017nanoscale}; 
Yarali \cite{yarali2017afm}; 
Aiyiti \cite{aiyiti2018nanoscale}; 
Gu \cite{gu2016jap}.}
\label{f_md_vs_exp}
\end{center}
\end{figure*}

In our previous simulations, we have only considered isotopically pure systems. As the experimental samples have not been isotopically purified, we include isotope disorder according to the isotope compositions compiled in Ref. \onlinecite{meija2016pac}. The calculated in-plane thermal conductivity of suspended monolayer is $110 \pm 4$ W m$^{-1}$ K$^{-1}$, which is nearly unchanged compared to that of isotopically pure MoS$_2$. This is similar to the finding by Li \textit{et al.} \cite{li2013apl} obtained by using the BTE method combined with DFT calculations. The small effect of isotope scattering on the thermal conductivity is expected as the mass mismatch between Mo and S is only slightly affected by the inclusion of naturally occurring isotopes.

Experimental measurements are available not only for single-layer MoS$_2$, but also for multi-layer and bulk MoS$_2$. We calculated the thermal conductivity of bulk MoS$_2$ (represented as six-layer MoS$_2$ with periodic boundary conditions in all directions) as well as two- to five-layer MoS$_2$, with isotope disorder included. The relevant results are presented in Table \ref{table:kappa} and also visualized in Fig. \ref{f_md_vs_exp}. The layer dependence of thermal conductivity is very similar to that obtained by Gu \textit{et al.} \cite{gu2016jap} based on BTE calculations; the thermal conductivity decreases with increasing layer number $n$ and saturates at $n=3$. It has been suggested that both the change of phonon dispersion and the thickness-induced anharmonicity associated with the breakdown of a mirror symmetry in single-layer MoS$_2$ are responsible for the reduction of thermal conductivity with increasing layer number \cite{gu2016jap}.

\begin{table}[htb]
\caption{Through-plane thermal conductivity $\kappa_z$ of bulk MoS$_2$ at $300$ K from experiments and our calculations using the HNEMD method with the REBO-LJ potential. Isotope scattering is considered in the HNEMD calculations. }
\begin{center}
\begin{tabular}{ l l l l l}
\hline
\hline
Ref.        & $\kappa_z$ (W m$^{-1}$ K$^{-1}$) \\
\hline
\cite{muratore2013apl}    &  $\sim 2.3$     \\
\cite{liu2014jap}         &  $2.0 \pm 0.3$  \\
\cite{zhu2016nc}         &  $2.0$  \\
\cite{jiang2017am}        &  $4.75 \pm 0.32$ \\
Here                      &  $2.2 \pm 0.2$  \\
\hline
\end{tabular}
\end{center}
\label{table:kz}
\end{table}

Apart from bulk MoS$_2$, our MD predicted values are consistently larger than experimental values, which however show large variations. Variations in the experimental results could be due to differences in the quality of each sample, measurement calibration, and the presence of thermal contact resistance. Indeed, some experimental samples \cite{sahoo2013jpcc,bae2017nanoscale,yarali2017afm} were grown by chemical vapor deposition (CVD), which are usually polycrystalline, consisting of grains separated by grain boundaries. It has been recently shown that dense grain boundaries can heavily reduce \cite{sledzinska2017acsami} the thermal conductivity of MoS$_2$. Even for exfoliated samples, there are defects and possibly rough edges \cite{aiyiti2018nanoscale} which can also reduce the thermal conductivity. Nevertheless, our MD results for single-layer and bi-layer MoS$_2$ are close to those measured by Zhang \textit{et al.} \cite{zhang2015acsami} on exfoliated samples. In the other limit of bulk MoS$_2$, our MD predicted value agree well with those measured on natural crystals (with high purity)  \cite{liu2014jap,zhu2016nc,jiang2017am}. Moreover, our calculated through-plane thermal conductivity $\kappa_z$ is also close to the experimental values, as can be seen from Table \ref{table:kz}. Based on all these comparisons, we conclude that the REBO-LJ potential has predictive power in terms of thermal transport properties in Mo-S systems and can be used for more spatially complex structures than pristine MoS$_2$, where the BTE-based method is less applicable. We leave these applications to future work.

\section{Summary and conclusions \label{section:conclusions}}

In summary, we have employed extensive classical MD simulations to study heat transport in single-layer, multi-layer, and bulk MoS$_2$. We considered three existing empirical many-body potentials for MoS$_2$ in the literature: the REBO-LJ potential \cite{liang2009prb,liang2012prb,stewart2013msmse}, the SW13 potential by Jiang \textit{et al.} \cite{jiang2013jap}, and the SW16 potential by Kandemir \textit{et al.} \cite{kandemir2016nt}. To calculate the thermal conductivity, we mainly used the highly efficient HNEMD method for many-body potentials \cite{fan2018submitted} and used the EMD and NEMD methods to check the consistency of our data. Most of the MD simulations were done using the efficient GPUMD code \cite{fan2013cpc,fan2017cpc,gpumd}, but the LAMMPS code \cite{plimpton1995jcp,lammps} was also used in some cases to double-check. For each empirical potential used, we have obtained consistent results between the different MD methods by using the GPUMD code. However, our results differ significantly from some previous studies in the literature. While we can understand the NEMD results by Ding \textit{et al.} \cite{ding2015nt} and Hong \textit{et al.} \cite{hong2016jpcc}, we failed, despite extensive efforts, to reproduce the EMD results by Jin \textit{et al.} \cite{jin2015sr}, Hong \textit{et al.} \cite{hong2016jpcc}, and Kandemir \textit{et al.} \cite{kandemir2016nt}. 

Based on our results for single-layer MoS$_2$, we found that both the SW13 and the SW16 potentials do not describe the phonon anharmonicity of MoS$_2$ properly: they lead to very slow convergence of the running thermal conductivity in the Green-Kubo relation, indicating the existence of phonon modes with very long relaxation times or very large mean free paths. In contrast, both the time scale and the thermal conductivity value from the REBO-LJ potential are very reasonable.

Finally, we took isotope scattering into account and evaluated the thermal conductivities of single-layer, multi-layer and bulk MoS$_2$ using the REBO-LJ potential and compared closely with predictions obtained from the BTE approach as well as available experimental data. We found that the thermal conductivity decreases with increasing layer number $n$ and saturates at $n=3$, which agrees with the prediction by Gu \textit{et al.} \cite{gu2016jap} from BTE calculations. Our predicted thermal conductivity values agree well with those measured on samples with relatively high quality \cite{zhang2015acsami,liu2014jap,zhu2016nc,jiang2017am}. We also compared the phonon dispersion curves calculated using the empirical potentials with available experimental data \cite{tornatzky2018arxiv}. From all these comparisons, we identify the REBO-LJ potential as a transferable empirical potential for MoS$_2$ that can be applied to study thermal transport properties of MoS$_2$ in more complicated situations such as systems with the presence of defects, grain boundaries or specifically engineered nanoscale features, where the BTE approach is less practical. Such applications will be considered in future studies. This work establishes a firm foundation for understanding heat transport properties of MoS$_2$ using MD simulations.

\begin{acknowledgments}
We thank Zhennan Kou for fruitful discussions. This work was supported in part by the National Natural Science Foundation of China (Grant Nos. 11404033 and 11502217) and in part by the Academy of Finland (Projects Nos. 286279 and 311058) and its Centre of Excellence program QTF (Project 312298). We acknowledge the computational resources provided by Aalto Science-IT project, Finland's IT Center for Science (CSC) and HPC of NWAFU. A.J.G. acknowledges support from the NDSEG Fellowship as well as Stanford University and the Stanford Research Computing Center for providing computational resources that contributed to these results. A.J.G. and E.P. acknowledge support from the Air Force Office of Scientific Research (AFOSR) grant FA9550-14-1-0251, the National Science Foundation (NSF) EFRI 2-DARE grant 1542883, and the Stanford SystemX Alliance. A.V.K. also thanks DFG for the support under Project No. KR 4866/2-1.
\end{acknowledgments}

\end{document}